\documentclass[fleqn,12pt,twoside]{article}
\usepackage{espcrc1}

\usepackage{graphicx}

\usepackage{amsmath,amssymb,bm}

\newcommand{\re}{\text{Re}}

\newcommand{\be}{\begin{equation}}
\newcommand{\ee}{\end{equation}}
\newcommand{\ba}{\begin{eqnarray}}
\newcommand{\ea}{\end{eqnarray}}

\title{Two-meson cloud contribution to the baryon
antidecuplet self-energy}

\author{T. Hyodo\address[RCNP]
                {Research Center for Nuclear Physics (RCNP),
                Ibaraki, Osaka 567-0047, Japan.},
	A. Hosaka\addressmark[RCNP],
	F.J.~Llanes-Estrada\address[Madrid1]
	        {Universidad Complutense de Madrid, Depto. F\'{\i}sica
                Te\'orica I, 28040 Madrid, Spain.},
	E.~Oset\address[Valencia]
	        {Departamento de F\'isica Te\'orica and IFIC,
                Centro Mixto Universidad de Valencia-CSIC,
		Institutos de Investigaci\'on de Paterna, Aptd. 22085, 46071
		Valencia, Spain.},
	J.~R.~Pel\'aez\address[Madrid2]
	        {Universidad Complutense de Madrid, Depto. F\'{\i}sica
                Te\'orica II, 28040 Madrid, Spain.} and
	M.~J.~Vicente~Vacas\addressmark[Valencia],
		}    
\begin{document}

\maketitle

\begin{abstract}
    We study the self-energy of the $SU(3)$ antidecuplet coming from
    two-meson virtual clouds.
    Assuming that the exotic $\Theta^+$ belongs to an antidecuplet
    representation with $N(1710)$ as nucleon partner,
    we derive effective Lagrangians
    that describe the decay of $N(1710)$ into $N\pi\pi$ with
    two pions in $s$- or $p$-wave.
    It is found that the self-energies for all members of the
    antidecuplet are attractive, and 
    the larger strangeness particle is more bound.
    From two-meson cloud, we obtain about  20\% of the empirical
    mass splitting between states with different strangeness.
\end{abstract}

\section{Introduction}

In recent years,
the study of the exotic pentaquark
baryons has been
one of the most exciting fields.
Evidence of exotic $\Theta^+$ was reported in Ref.~\cite{Nakano:2003qx}
and a signal of another exotic state $\Xi^{--}$
was subsequently observed~\cite{Alt:2003vb}, though the 
experimental confirmation of the latter
is somehow controversial.
In hadron spectroscopy, the Gell-Mann--Okubo[GMO] formula has been
successfully applied to describe the mass splitting of particles 
within an $SU(3)$ multiplet.
The smallest multiplet containing $\Theta^+$ and $\Xi^{--}$ is the
antidecuplet ($\overline{\bm{10}}$), and equal mass splitting is obtained 
from the GMO rule.
Here we would like to study the two-meson cloud effect
to the baryon antidecuplet, which would contribute to mass splitting
in addition to the GMO formula.

The study of the two-meson cloud effect is motivated by the attempts of
constructing the $\Theta^+$ as a $K\pi N$ bound
state~\cite{Bicudo:2003rw,Kishimoto:2003xy,Llanes-Estrada:2003us,Bicudo:2004cm},
in which an attractive interaction
is found for the $K\pi N$ states of  $J^P=1/2^+$.
Although the found attraction is not enough to bind the three
body system, such a configuration is naturally expected
in the $\Theta^+$ structure, by observing that
the $K\pi N$ mass is only 30 MeV above
the $\Theta^+$ state.

In the present work, assuming
that the $N(1710)$ has a large
antidecuplet component,
we construct flavor $SU(3)$ effective interaction Lagrangians,
which account for the decay modes of the $N(1710)$ into $N
\pi\pi$.
Using these interaction Lagrangians,
we calculate the self-energy of the antidecuplet.
Details of the study can be found in Ref.~\cite{Hosaka:2004mv}.

\section{Formulation}\label{sec:Lag}

The interaction Lagrangians are constrained to be 
$SU(3)$ symmetric.
The process we are considering is
$ \bm{8}_M + \bm{8}_M + \bm{8}_B \to \bar{\bm{10}}_P $,
where we denote the octet baryon, meson and antidecuplet
baryon as $\bm{8}_{B}$, $\bm{8}_{M}$ and $\bar{\bm{10}}_P$,
respectively.
In order to construct a singlet from the product of 
$\bm{8}_{M}$, $\bm{8}_{M}$, $\bm{8}_{B}$, and $\bar{\bm{10}}_P$,
there are four combinations, in which
two $\bm{8}_{M}$ mesons are combined into
$\bm{8}^s_{MM}$, $\bm{8}^a_{MM}$, $\bm{10}_{MM}$ and
$\bm{27}_{MM}$. However, two of them ($\bm{8}^a_{MM}$, $\bm{10}_{MM}$)
are identically zero, due to additional symmetry under
exchange of two mesons.
Hence, without using derivatives,
we can construct the following effective Lagrangians:
\begin{align}
    \mathcal{L}^{8s}
    =&\frac{g^{8s}}{2f}
    \bar{P}_{ijk}\epsilon^{lmk}
    \phi_{l}{}^{a}\phi_{a}{}^{i} B_{m}{}^{j}+ h.c. \ ,
    \label{eq:8sLag} \\
    \mathcal{L}^{27}
    =&\frac{g^{27}}{2f}
    \Bigl[4\bar{P}_{ijk}\epsilon^{lbk}
    \phi_{l}{}^{i} \phi_{a}{}^{j}B_{b}{}^{a}
    -\frac{4}{5}\bar{P}_{ijk}\epsilon^{lbk}
    \phi_{l}{}^{a} \phi_{a}{}^{j}
    B_{b}{}^{i}
    \Bigr] + h.c.  \ ,
    \label{eq:27Lag}
\end{align}
where $P$, $B$ and $\phi$ are the baryon antidecuplet, baryon octet
and meson octet fields, respectively.
We have included a factor
$1/2f$ to make $g^{8s}$ and $g^{27}$ dimensionless 
($f=93$ MeV is the pion decay constant).
In the low momentum expansion, the above Lagrangians are the two
lowest ones. In practice, however, they are not sufficient
to account for the experimental decay of $N(1710)$
into $N\pi\pi$($p$-wave).
In order to reproduce such decay mode,
we introduce a Lagrangian with one derivative:
\begin{equation}
    \mathcal{L}^{8a}
    =i\frac{g^{8a}}{4f^2}
    \bar{P}_{ijk}\epsilon^{lmk}\gamma^{\mu}
    (\partial_{\mu}\phi_{l}{}^{a} \phi_{a}{}^{i}
    - \phi_{l}{}^{a}\partial_{\mu} \phi_{a}{}^{i})B_{m}{}^{j} + h.c.
    \ .
    \label{eq:8aLag}
\end{equation}
We will address other possible interaction 
Lagrangians~\cite{Hosaka:2004mv} later on.

\begin{figure}[tbp]
    \centering
    \includegraphics[width=6.5cm,clip]{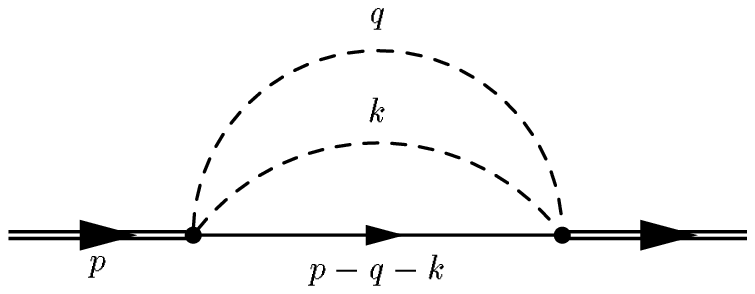}
    \includegraphics[width=6.5cm,clip]{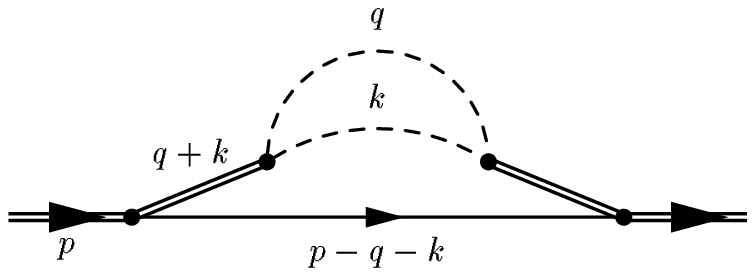}
    \vspace{-1cm}
    \caption{
    Diagrams for self-energy of baryon antidecuplet due to
    two-meson cloud.
    Right : inclusion of vector meson propagator.}
    \label{fig:loop}
\end{figure}

The antidecuplet self-energies are given by
\begin{equation}
    \Sigma^{(j)}_{P}(p^0)
    = \sum_{B,m_1,m_2} \left( F^{(j)}
    C^{(j)}_{P,B,m_1,m_2} \right) 
    I^{(j)}(p^0;B,m_1,m_2) 
    \left(F^{(j)} C^{(j)}_{P,B,m_1,m_2} \right) \ ,
    \label{eq:selfenergy}
\end{equation}
where the index $j$ labels the interaction
Lagrangians, $p^0$ is the energy of the antidecuplet baryon,
$F^{(j)}$ are
coupling constants appearing in the Lagrangian,
and $C^{(j)}_{P,B,m_1,m_2}$ are $SU(3)$ coefficients which are
compiled in the Appendix of Ref.~\cite{Hosaka:2004mv}.
The function $I^{(j)}(p^0;B,m_1,m_2)$
is the two-loop integral with two mesons and one baryon
(Fig.~\ref{fig:loop}, left):
\begin{equation}
    \begin{split}
	&I^{(j)}(p^0;B,m_1,m_2) \\
	=& -
	\int \frac{d^4k}{(2\pi)^4}
	\int \frac{d^4q}{(2\pi)^4}
	|t^{(j)}|^2
	\frac{1}{k^2-m_1^2+i\epsilon}
	\frac{1}{q^2-m_2^2+i\epsilon}
	 \frac{M}{E} 
	\frac{1}{p^0-k^0-q^0-E+i\epsilon} \ ,
    \end{split}
    \label{eq:loop2meson}
\end{equation}
where $t^{(j)}$ are the amplitudes derived from the Lagrangian $j$,
$M$ and $m_i$ are the masses of a baryon
and mesons, $E$ is the energy of the intermediate baryon.
The real part of this integral is 
cut off with a three momentum $\Lambda$
in the range 700-800 MeV.
The imaginary part of the diagram 
provides the decay width.

It is known that $N(1710)\to N\pi\pi(p$-wave) occurs through 
the $N\rho$ decay, therefore, we
improve the contact interaction of the $\mathcal{L}^{8a}$ to account 
for the vector meson propagator (Fig.~\ref{fig:loop}, right), 
including the factor $m_v^2/[(q+k)^2-m_v^2]$
in each $P\to BMM$ vertex.

\section{Numerical results}\label{sec:Results}

Here we show the results with $\mathcal{L}^{8s}$
and $\mathcal{L}^{8a}$.
$\mathcal{L}^{27}$ and other
possible Lagrangians will be addressed later.
The parameters $g^{8s}$ and $g^{8a}$ are fixed so as to
to obtain the partial decay widths
of the $N(1710)$ to $N\pi\pi(s$-wave, isoscalar) and 
$N\rho\to N\pi\pi(p$-wave, isovector)
respectively. 
The central values in the PDG~\cite{Eidelman:2004wy} are
25 and 15 MeV,
which correspond to
$g^{8s} = 1.9$ and $g^{8a} = 0.32$, respectively.
We take an average value of $p^0=1700$ MeV as input.
The qualitative trend of the result does not depend on the $p^0$,
but the magnitude of the self-energy is changed~\cite{Hosaka:2004mv}.

\begin{figure}[tbp]
    \centering
    \includegraphics[width=7cm,clip]{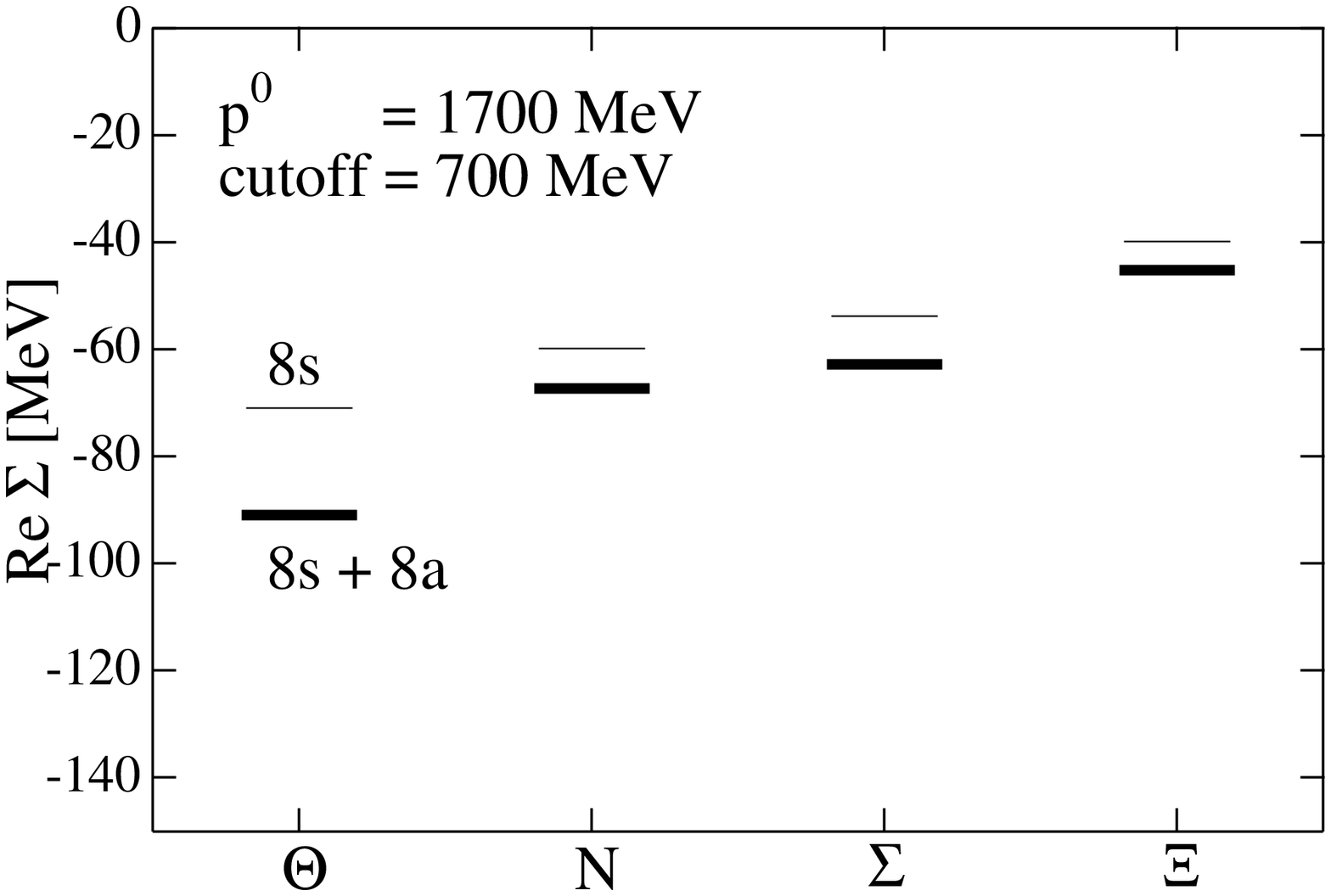}
    \includegraphics[width=7cm,clip]{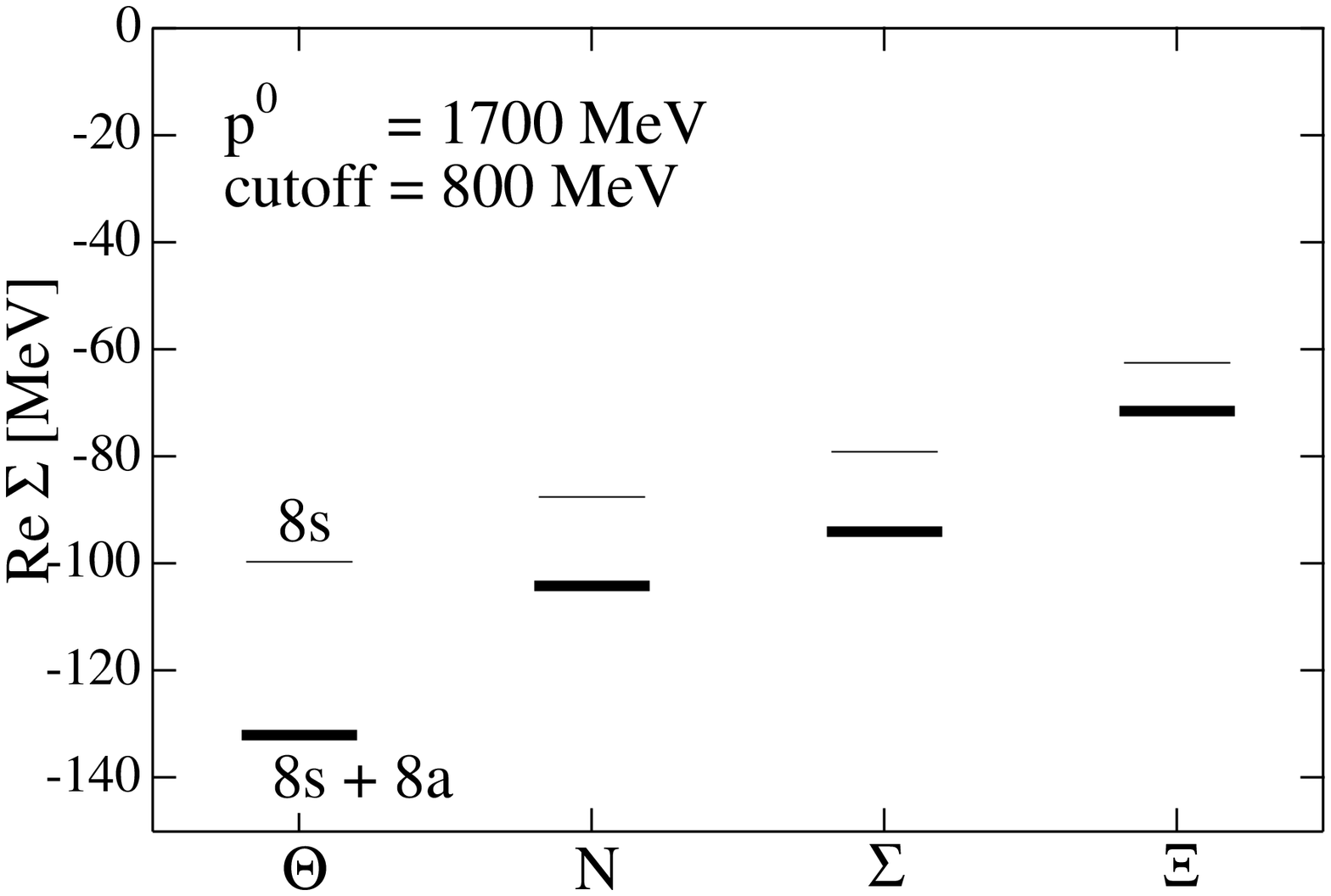}
    \vspace{-1cm}
    \caption{
    Mass shifts of baryon antidecuplet $(\re \Sigma_P)$ 
    with $p^0 = 1700$ MeV.}
    \label{fig:totresult}
\end{figure}

In Fig.~\ref{fig:totresult}
we show the real parts of the self-energies for the
contributions from $\mathcal{L}^{8s}$ 
and total contributions of  $\mathcal{L}^{8a}$
and  $\mathcal{L}^{8s}$, with cutoffs  
$700$ and $800$ MeV.
We see that all the self-energies are attractive,
and that the interaction is more attractive  the larger
the strangeness.
$\mathcal{L}^{8s}$ provides more binding than 
$\mathcal{L}^{8a}$ for the same cutoff.
The splitting between the $\Theta_{\bar{10}}$ and $\Xi_{\bar{10}}$ states
is about 45-60 MeV depending on the cutoff.
Since the experimental splitting of the $\Theta(1540)$ and $\Xi(1860)$
is 320 MeV,  the two-meson cloud
provides 20\% of the total splitting.
This should be compared to 60\% naturally provided by the mass of
the constituent strange quarks, which would leave about 20\%
more for the effects of quark correlations.

The partial decay widths are shown in Table~\ref{tbl:decay}.
We have taken the observed masses
$M_{N_{\bar{10}}} = 1710$, $M_{\Sigma_{\bar{10}}} = 1770$
and $M_{\Xi_{\bar{10}}} = 1860$ MeV as $p^0$,
because the phase space is essential for the
decay width.
We can see that the widths are not very large for all channels.
Indeed, $\Sigma(1770)$ and $\Xi(1860)$ would have widths
for three body channels of about 24 and 2 MeV,
which are compatible with the experimental total width 
of about 70 and 18
MeV, respectively~\cite{Alt:2003vb,Eidelman:2004wy}.
Detailed information of the partial decay widths of
three body channels will give us more understanding
of the interaction Lagrangian.

\begin{table}[btp]
    \caption{Partial decay widths for the allowed channels and
    total width for any $BMM$ channel, at the
    masses of the antidecuplet members. All values are listed
    in units of MeV.}
    \centering
    \renewcommand{\tabcolsep}{2pc} 
    \renewcommand{\arraystretch}{1.2} 
    \begin{tabular}{llll}\hline
    Decay  widths [MeV] & $\Gamma^{(8s)}$ & $\Gamma^{(8a)}$
    & $\Gamma^{tot}_{BMM}$ \\ \hline
    $N(1710)\to N\pi\pi$ (inputs) & 25 & 15 & 40 \\
    $N(1710)\to N\eta\pi$ & \phantom{ }0.58 & - & \\ 
    $\Sigma(1770)\to N\bar{K}\pi$ & \phantom{ }4.7 
    & \phantom{ }6.0 & 24 \\
    $\Sigma(1770)\to \Sigma\pi\pi$ & 10 & \phantom{ }0.62  &\\
    $\Sigma(1770)\to \Lambda\pi\pi$ & - & \phantom{ }2.9 & \\
    $\Xi(1860)\to \Sigma \bar{K}\pi$ & \phantom{ }0.57 
    & \phantom{ }0.46 &  \phantom{ }2.1 \\
    $\Xi(1860)\to \Xi \pi \pi$ & - & \phantom{ }1.1 & \\
    \hline
    \end{tabular}
    \label{tbl:decay}
\end{table}

In Ref~\cite{Hosaka:2004mv},
apart from the above Lagrangians and a ${\cal L}^{27}$ term,
we also considered the leading order Lagrangians in a chiral expansion
as dictated by QCD. There are two terms:
one chirally symmetric and a mass term.  The latter has to be tiny
since it violates SU(3). The former gives similar results to
${\cal L}^{8s}$.
The role of chiral symmetry is thus small and the use of
${\cal L}^{8s}$ is justified.

\section{Summary}\label{sec:Summary}

We study the self-energy of the baryon antidecuplet due to the 
two-meson cloud.
The assumptions made throughout the paper and the uncertainties in
the experimental input make the nature of our analysis qualitative.
However, in all different cases studied, 
the two-meson cloud mechanism leads to the following conclusions:
An attractive self-energy is obtained for all members of the
antidecuplet.
The two-meson cloud contributes to the mass splitting between
antidecuplet members about 20\% of the empirical one.
These observations are consistent with the previous attempts to describe 
the $\Theta^+$ as a $K\pi N$ 
state~\cite{Bicudo:2003rw,Kishimoto:2003xy,Llanes-Estrada:2003us,Bicudo:2004cm},
and the magnitude of 20\% is also in agreement quantitatively with 
the strength of attraction~\cite{Llanes-Estrada:2003us}.
The role played by the two-meson cloud is therefore of
relevance for a precise understanding of the nature
of the $\Theta^+$  and the antidecuplet.

\end{document}